\documentclass[12pt]{article}
\usepackage{mathrsfs}
\usepackage{amsmath}
\usepackage{amsfonts}
\usepackage{amssymb}
\usepackage{latexsym}
\usepackage{graphicx}
\usepackage[english]{babel}
\usepackage{epsfig}
\def\newhorizon{\gamma^{2/3}u_{T} }
\def\tension{2\pi \alpha'}
\def\be{\begin{equation}}
\def\ee{\end{equation}}
\newcommand{\bea}{\begin{eqnarray}}
\newcommand{\eea}{\end{eqnarray}}
\newcommand{\non}{\nonumber \\}
\newcommand{\CR}{\non\cr}
\def\Im{{\rm Im}}
\def\t{{\hat t}}

\def\y{{\hat y}}
\def\Y{{\hat Y}}
\def\u{{\hat u}}

\def\w{{\hat \omega}}
\def\M{{\hat{M}}}
\def\n{{\hat n}}
\def\R{{\hat R}}
\def\q{{\hat q}}
\def\x{\hat{x}}
\def\E{{\hat E}}
\def\T{{\rm T}}
\def\const{\rm const}

\begin{document}

\title{Transverse momentum broadening of heavy quark and gluon energy loss in Sakai-Sugimoto model}

\author{Yi Pang}


\hfill \hbox{{USTC-ICTS-08-05}} \vspace*{5.0ex}
\begin{center}

{\Large \bf Transverse momentum broadening of heavy quark and gluon
   energy loss in Sakai-Sugimoto model}
\end{center}
\vspace{10pt}
\begin{center}
{\Large Yi Pang$^{2,1}$}
\end{center}


 \vspace{20pt}

\begin{center}

               {\em
    $^{1}$ Interdisciplinary Center of Theoretical Studies, USTC,\\
    Hefei, Anhui 230026, P.R.China\\
    $^{2}$ Institute of Theoretical Physics, CAS, Beijing 100080,
    P.R.China}

\end{center}

\vspace*{5.0ex} \centerline{\tt yipang@itp.ac.cn} \vspace*{5.0ex}

\begin{center}
\begin{abstract}

 In this paper, we calculate the transverse momentum diffusion coefficient $\kappa_{T}$ of heavy quark and gluon penetration length in the
 deconfinement phase of Sakai-Sugimoto model, which is known as a holographic dual of large
 $\emph{N}_{c}$ QCD. We find that for the heavy quark moving through
 the thermal plasma with a constant velocity $v<1$, the transverse momentum diffusion coefficient
 $\kappa_{T}\propto\lambda\gamma^{\frac{1}{3}}\emph{T}^{4}/\emph{T}_{d}$,
 and the gluon penetration length $\triangle x\propto\hat{\emph{E}}^{\frac{2}{5}}$. These results are different from those calculated in $\mathcal{N}=4$
 super-Yang-Mills theory, which are
$\kappa_{T}\propto\lambda\gamma^{\frac{1}{2}}\emph{T}^{3}$ and
 $\triangle x\propto\hat{\emph{E}}^{\frac{1}{3}}$, respectively. In the
 high energy limit, the difference between the two pairs of results
 should be evident, so we hope that the future LHC experiments can
 tell us which model is more closely related to the realistic strongly coupled QCD at finite temperature.

 \end{abstract}

\end{center}
\vspace{60pt}

\newpage
 \section{Motivation}
\label{S:Introduction}

 The experimental relativistic heavy ion collisions have produced
 much evidence signalling that Quark Gluon Plasma (QGP) has
 been formed at the Relativistic Heavy Ion Collision (RHIC) \cite{ex-Adms05,
 ex-Adcox04}. One piece of strong evidence is that the production of the
 high-$p_T$ particles is suppressed \cite{ex-Adler05,ex-Bielcik05}. To explain this
 phenomenon within the framework of QCD is difficult, because recently, researchers have found that QGP is a strongly coupled fluid.
 In the framework of AdS/CFT \cite{Maldacena:1997re}, one can solve
 problems in strongly coupled gauge theories by considering the
 corresponding problems in dual weak coupled gravity theories. So,
 many people try to solve these problems in QGP, by transferring them into a gauge theory which has a gravity dual and can mimic
 QCD to some extent. Along this way, H. Liu, K. Rajagopal and U.
 Wiedemann define the jet
 quenching parameter $\hat{q}$, via a light-like Wilson loop \cite{jetquench-LRW1}. $\q$ is the transverse momentum squared
 transferred from medium to either the initial parton or the radiated gluon, it is related to the average medium-induced
 parton energy loss by BDMPS formalism \cite{BDMPS}. Meanwhile, J.
 Casalderrey-Solana, D. Teaney and S. Gubser prefer to use the
 transverse momentum coefficient $\kappa_{T}$
\cite{Solana-Teaney1,Solana-Teaney2,Gubser1}. Although the two
groups adopt different
 parameters, the calculations are both carried out in the same
 background, which is the $AdS_{5}$-Schwarzschild
 space-time. Subsequent work \cite{jetquench-LinMatsuo,
 jetquench-AvarmisSfetsos,
 jetquench-ArmestoEdelsteinMas} includes computing $\q$ in
 backgrounds with non-zero chemical potential. Above all, the background
 metrics they use usually involve an asymptotically $AdS_{5}$
 component, since the gravity theory in $AdS_{5}$ is dual to the $\mathcal{N}=4$ SYM,
  their results actually apply to $\mathcal{N}=4$
 SYM. However $\mathcal{N}=4$ SYM is not the same as QCD, so it is
 problematic whether or not their results really capture some features of
 QCD, if it does, then these features should also appear in
 other models approximating QCD and having gauge/string duality,
 since all these models belong to one framework.

 Fortunately, in paper \cite{SSmodel}, Sakai and Sugimoto provide us such an new
 model, which we call S-S
 model in this paper. This model is a holographic dual of four-dimensional, large $\emph{N}_c$ QCD in the low
 energy regime. In the high energy regime, the gauge theory in S-S model shows some
 differences from QCD, such as K-K modes.
 A lot of papers \cite{SSmodel-phase} have been done
 on this model, in which they recover some features similar to realistic QCD,
 such as confinement-deconfinement phase transition and chiral
 symmetry breaking-restoration phase transition.\footnote{The low as well as high spin mesons and their motions through
 QGP were studied in \cite{meson}.}
 The calculations about screening length and jet
 quenching parameter $\hat{q}$ in this model \cite{drag-Talavera,Yi-hongGao} have been carried out. But the transverse
 momentum diffusion coefficient $\kappa_{T}$ has not been obtained. To give a complete
 comparison between above two models, we calculate $\kappa_{T}$ in
 S-S model. During the preparation of this paper, S. Gubser et al \cite{Gubser2} put
 forward a new approach to estimate the jet quenching parameter $\hat{q}$
 by considering the gluon energy loss in the thermal plasma of strongly coupled $\mathcal{N}=4$ super-Yang-Mills theory.
 Using this new method, we estimate $\hat{q}$ in S-S model. If we
 did not try this new way in S-S model, the comparison
 between the two models is still incomplete.

 This paper is organized as follows. In
 Section 2 we give a brief review of S-S model. In Section 3,
 after a short review of momentum diffusion constant in
 $\mathcal{N}=4$ SYM, we calculate the same transport coefficient in
 S-S model. In Section 4, we compute the lower and upper bound of
 gluon penetration length in S-S model and prepare to estimate $\q$. In Section 5, we use
 the results of the previous two sections to perform a quantitative analysis.

\section{A brief review of S-S model}
\label{S-S review} In \cite{SSmodel}, Sakai and Sugimoto present a
holographic dual of four-dimensional, large $N_c$ QCD. This model is
constructed by placing $N_f$ probe D8-$\overline{D8}$ into $N_c$ D4
brane background($N_f \ll N_c$ ), where supersymmetry is completely
broken by compactifying the $N_c$ D4 branes on a circle of radius
$R$ with anti-periodic boundary conditions for fermions
\cite{Witten}. At low energy, the D4/D8/$\overline{D8}$ system
yields a $U(N_c)$ gauge theory with fermions, and there is also a
$U(N_f)_L\times U(N_f)_R$ chiral symmetry. Being well studied in
\cite{SSmodel-phase}, this model contains confinement-deconfinement
phase transition, the critical temperature is $\emph{T}_{d} = 1
/2\pi R$.

 When the system arrives at a temperature $\emph{T }<\emph{T}_{d}$, the dual gauge theory of S-S model is
in the confined phase, we should use the following background to
describe it
\bea\label{SSmodel} ds^2&=&\left( \frac{u}{R_{D4}}
\right)^{3/2}\left [- dt^2 +\delta_{ij}dx^i dx^j + f(u) dx_4^2
\right ] +\left( \frac{R_{D4}}{u} \right)^{3/2} \left [
\frac{du^2}{f(u)} + u^2 d\Omega_4^2 \right ], \CR F_{(4)}&=&
\frac{2\pi N_c}{V_4}\epsilon_4, \quad e^\phi = g_s\left(
\frac{u}{R_{D4}} \right)^{3/4},
\quad
 R_{D4}^3 \equiv \pi g_s N_c l_s^3,\quad
f(u)\equiv 1-\left( \frac{u_\Lambda}{u} \right)^3,
\eea
where $t$ is the time direction and $x^i$ ($i=1,2,3$) are the
 uncompactified world-volume coordinates of the D4 branes, $x_4$ is a
 compactified direction of the D4-brane world-volume which is transverse to
 the probe D8 branes, the volume of the unit four-sphere $\Omega_4$ is
 denoted by $V_4$ and the corresponding volume form by $\epsilon_4$,
 $l_s$ is the string length and finally $g_s$ is a parameter related
 to the string coupling. The submanifold of the background spanned by
 $x_4$ and $u$ has the topology of a cigar. The tip of the cigar is non-singular if and
 only if the periodicity of $x_4$ is
\be
\delta x_4 = \frac{4\pi}{3}\left(
 \frac{R_{D4}^3}{u_\Lambda}\right)^{1/2} = 2\pi R
.\ee

 When $\emph{T }>\emph{T}_{d}$, deconfinement happens, we should use another background to depict the dual gauge theory,
\bea \label{deconfinement} ds^2&=&\left( \frac{u}{R_{D4}}
\right)^{3/2}\left [-f(u) dt^2+ \delta_{ij}dx^{i}dx^j + dx_4^2\right
] +\left( \frac{R_{D4}}{u} \right)^{3/2} \left [ u^2 d\Omega_4^2 +
\frac{du^2}{f(u)} \right ], \label{unflavmetr}\CR F_{(4)}&=&
\frac{2\pi N_c}{V_4}\epsilon_4, \quad e^\phi = g_s\left(
\frac{u}{R_{D4}} \right)^{3/4},\quad
 R_{D4}^3 \equiv \pi g_s N_c l_s^3,\quad
f(u)\equiv 1-\left( \frac{u_T}{u} \right)^3. \eea
This background involves a black hole. The Euclidean time direction
$t_{E}$ now shrinks to zero size at the minimal value of $u$,
$u=u_T$. In order to avoid singularity, the Euclidean time direction
must have a period of
\be
 \delta t_{E} = \frac{4\pi}{3}\left( \frac{R_{D4}^3}{u_T}
\right)^{1/2} = \beta .\ee

QGP is the deconfined phase of QCD, in the following, we focus on
the deconfined phase.

In the deconfined phase, there exist two kinds of configurations of
the probe D8 and $\overline{D8}$ branes. The one depicted in Figure
(1a) signals the breaking of chiral symmetry, and the other
indicates the restoration of chiral symmetry.

We should remind the reader that, in this paper, the probe branes
only serve as the place where the string hangs. Our calculations are
independent of the detailed brane configurations.
\begin{figure}[ht]
\begin{center}
\vspace{0ex}
\includegraphics[width=.25\textwidth]{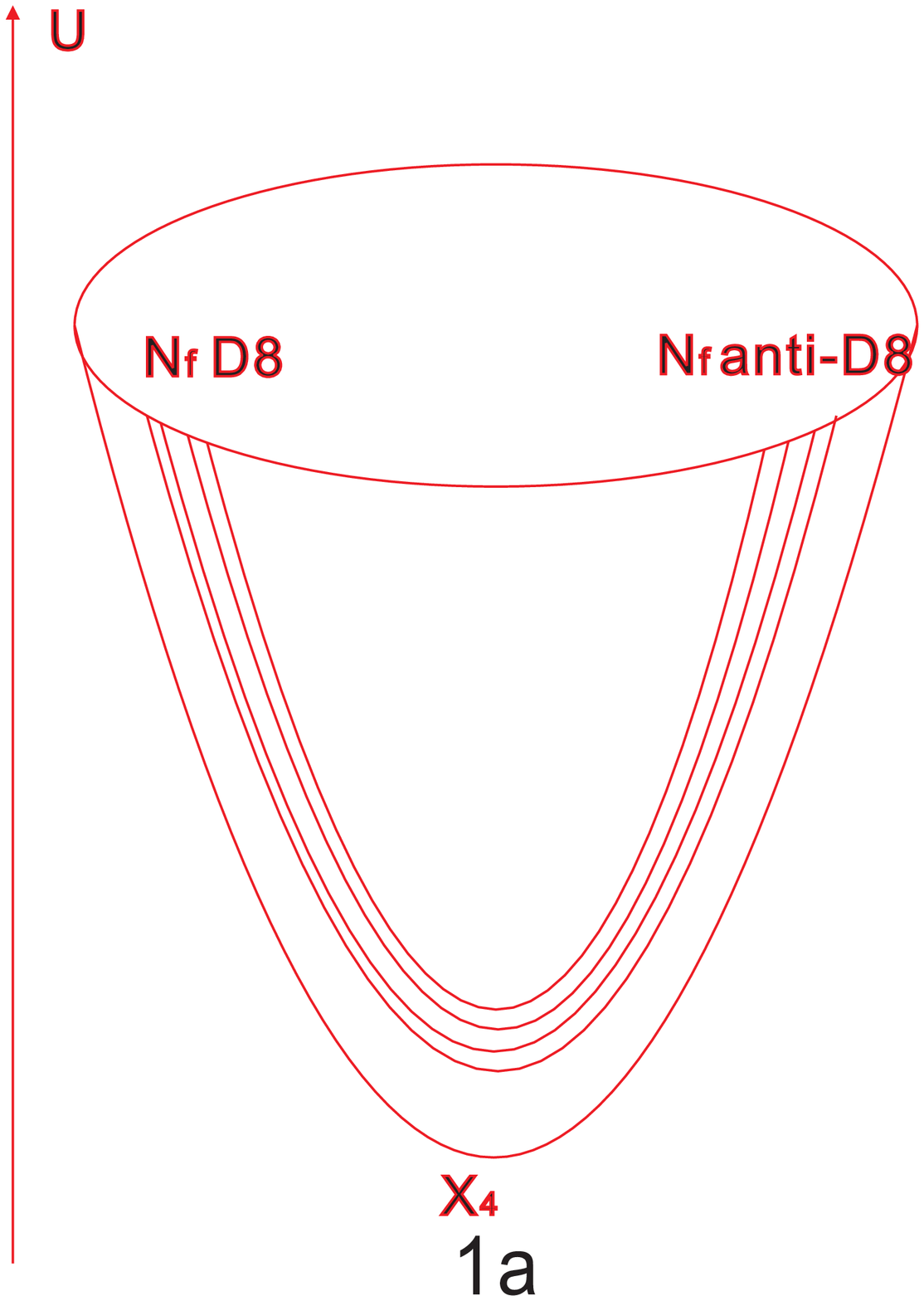}
\includegraphics[width=.25\textwidth]{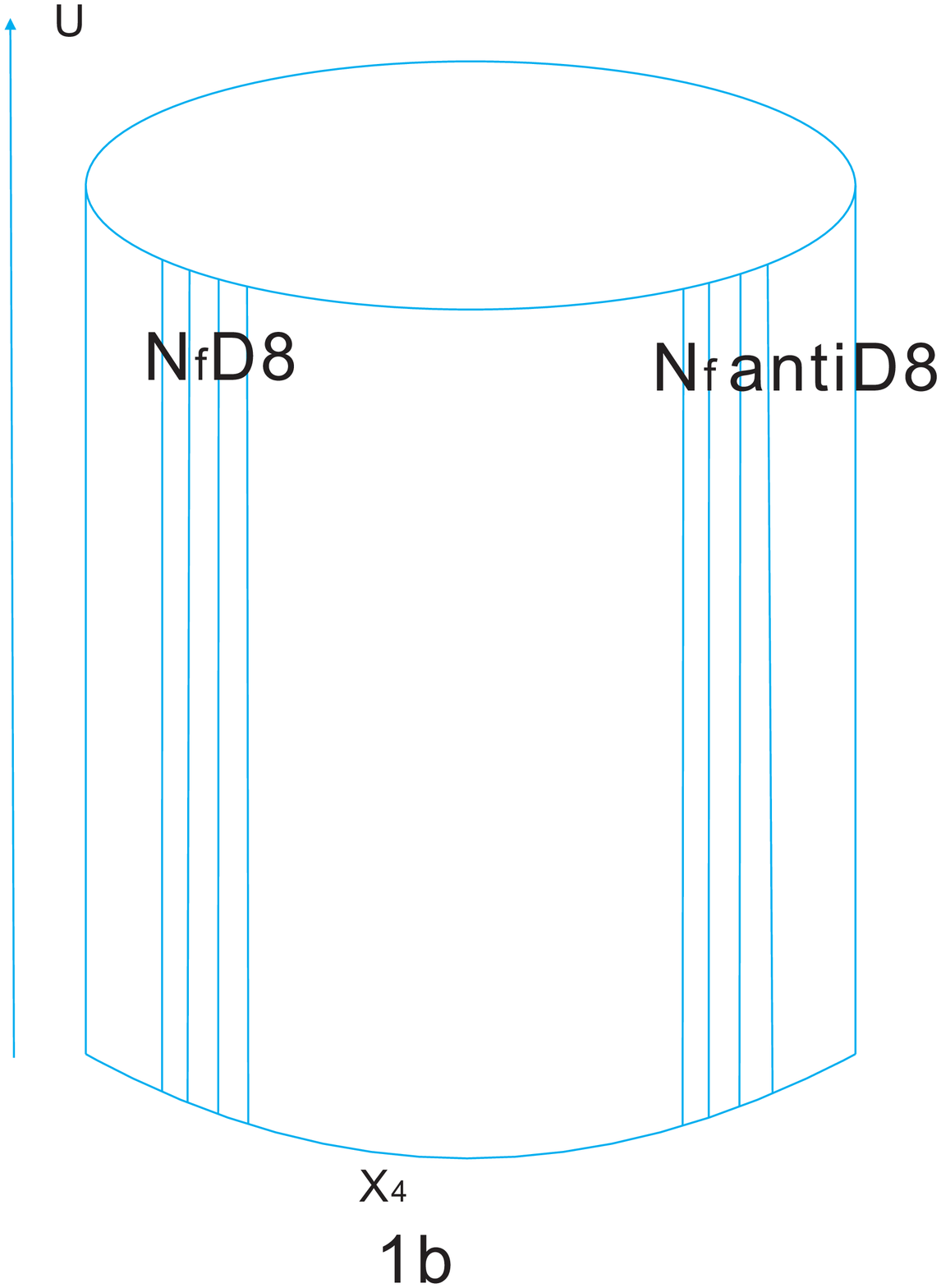}
\end{center}
\caption{The dominant configurations of the D8 and anti-D8 probe
branes in Sakai-Sugimoto model in deconfinement phase. (1a)
indicates the chiral symmetry is broken, and (1b) signifies the
chiral symmetry is restored. If D8 and $\overline{D8}$ are separated
by a distance L at infinity, for $T>0.154/L$, (1b) is the dominant
phase, for $T_{d}<T<0.154/L$, (1a) is the dominant phase}
\end{figure}

\section{Calculation of the momentum diffusion coefficient}

\subsection{Preliminaries}

In \cite{Solana-Teaney2,Gubser1} the authors obtain the transverse
momentum diffusion coefficient in the following way. Firstly, by
analogy to classical theory for Brownian motion, they propose that
\be\label{mdc} \kappa_T = \frac{1}{2}\int_{-\infty}^\infty dt \,
\langle F(t)F(0) +  F(t) F(0) \rangle ,\ee
 where $F(t)$ is the transverse stochastic force acting on the probe quark. In
 $\mathcal{N}=4$ SYM, it takes the form
 \be\label{force} F(t) = \int d^{3}\vec{x}
 Q^{\dag}(t,\vec{x})T^{a}Q(t,\vec{x})E^{a},\ee
 where $E^{a}$ is the field strength supplied by vector gauge fields and six scalar
 fields. We can define the Wightman correlation function and the
 Feynman correlation function about $F(t)$, they are
 \be G(t) =\frac{1}{2}\langle F(t)F(0) + F(t) F(0) \rangle,\ee
 \be G_{F}(t)= -i\langle {\T F(t)F(0)}\rangle.\ee
 And there is a relation between
 them in frequency space, \be G(\omega)= - \Im G_{F}(\omega).\ee
 So, Eq. (\ref{mdc}) can be changed into \be \kappa_T = \lim_{\omega \to 0}
 G(\omega) = -  \lim_{\omega \to 0} \Im G_{F}(\omega).\ee
 The $\langle~~~\rangle$ denotes an average in the states which are
 composed of SYM and a moving quark. Using Wigner distribution
 function in QCD kinetic theory, the generating functional of the above correlation
 functions can be written as the VEV of a Wilson loop. This loop is a
 closed contour in the complex time plane, specially, its $x$
 component should satisfy $x = x_{0} + v (t_{c}-t_{c_{0}})$ and $y$
 component is equal to $\delta$$y_1(t_c)$ when $t_c$ lies on the real
 time axis, $\delta$$y_2(t_c)$ when $t_c$ lies below the real time
 axis. This choice is determined by that the probe quark is traveling in $x$ direction with velocity
 $v$, and $\delta$$y_{1}(t)$, $\delta$$y_{2}(t)$ are the fluctuations of quark's displacement in transverse direction
 acting as external sources coupling to the transverse stochastic force.

The authors of paper \cite{Solana-Teaney1,Solana-Teaney2,Gubser1}
evaluate the VEV of the Wilson loop via AdS/CFT correspondence in
$AdS_{5}$-Schwarzschild background, by finding out a string's
classical action, requiring that the boundary of string's
world-sheet is the Wilson loop. This is to say
\be\label{gauge-string} \frac{1}{e^{iS_{NG}[0,0]}} e^{iS_{NG}[\delta
y_1, \delta y_2]} =  \frac{1}{\langle W[0,0] \rangle} \, \langle
W[\delta y_1, \delta y_2] \rangle.\ee
 The closed time contour corresponds to the two boundaries of global $AdS_{5}$-Schwarzschild
space and the string stretches between the two boundaries. Finally,
they obtain \be{} \kappa_T = \sqrt{\gamma \lambda} T^3 \pi \, , \ee
when $v \rightarrow 0, \gamma\rightarrow 1$, $\kappa_T$ and the drag
coefficient \cite{Herzog to Yaffe,Gubser3} satisfy the Einstein
relation.

\subsection{Calculation of the $\kappa_{T}$ in S-S model}
In S-S model, $E^{a}$ appearing in the stochastic force term
(\ref{force})  should also include the contribution from K-K modes
with mass of the order of QGP temperature. We will use the
deconfinement phase background (\ref{deconfinement}). For
convenience, we use $u_{T}$ to scale dimensional coordinates and
other parameter.
 \bea \label{metric} ds^2&=u_{T}^{2}&\left(
\frac{u}{R} \right)^{3/2}\left [ -f(u) dt^2+ \delta_{ij}dx^{i}dx^j +
dx_4^2\right ] +\left( \frac{R}{u} \right)^{3/2} \left [ u^2
d\Omega_4^2 + \frac{du^2}{f(u)} \right ], \eea
 where $R = \frac{R_{D_{4}}}{u_{T}}$, $f(u) = 1-\frac{1}{u^{3}}$,
 and now all the coordinates are dimensionless. Since this
 background is spherically symmetric , to define its Kruskal
 coordinates is a routine. They are
\be U = e^{-2\nu_{+}},\quad V =e^{2\nu_{-}},\ee
 where \be \nu_{+} \equiv t + z_{*},\quad \nu_{-} = t - z_{*},\ee
 and \be z_{*} = \int^{u}\frac{du}{f(u)(\frac{u}{R})^{3/2}}~~~.\ee

But we only need the near horizon behavior of the Kruskal
coordinates. In the near horizon limit, the metric becomes \be
 ds^2\sim 3u_{T}^{2}R^{-3/2}\left[ -(1-\frac{1}{u}) dt^2 + \frac{R^{3}du^2}{9(1-\frac{1}{u})}
 \right]
 + u_{T}^{2}
 R^{-3/2}\left [\delta_{ij}dx^{i}dx^j + dx_4^2+ R^{3}d\Omega_4^2  \right ]
 .\ee
 If we define
 \be \rho = \frac{R^{3/2}}{3}u,\quad 2M = \frac{R^{3/2}}{3}~~~,\ee
 then this metric looks like Schwarzschild metric
\be
 ds^2\sim 3u_{T}^{2}R^{-3/2}\left [ -(1-\frac{2M}{\rho}) dt^2 + \frac{d\rho^2}{1-2M/\rho}
 \right ]
 + u_{T}^{2}
  R^{-3/2}\left [\delta_{ij}dx^{i}dx^j + dx_4^2+ R^{3}d\Omega_4^2\right]~.
 \ee
 The near horizon Kruskal coordinates are the same as in the
 Schwarzschild metric,
\be U = -4Me^{-(t-r_{*})/4M},\quad V = 4Me^{(t-r_{*})/4M}~,\ee
 where \be r^{*} = \rho + 2M \ln |\rho/2M - 1|.\ee
 In background (\ref{metric}) the string configuration with a constant velocity $v$ in $x^{1}$ direction
 is as \cite{drag-Talavera}
 \be\label{trailing} x^{1}(t,u) = vt + \frac{v R^{3/2}}{3} \ln
 \frac{|u-1|}{\sqrt{u^{2}+u+1}} - \frac{v
 R^{3/2}}{\sqrt{3}}\arctan\frac{2u+1}{\sqrt{3}}~.\ee
with other coordinates kept constant, where we have chosen the $t$
and $u$ coordinates to parameterize the string world-sheet. So the
string configuration with a perturbation $\delta y(t,u)$ in the
$x^{2}$ direction should be \be x^{\mu} = (t, x^{1}(t,u), \delta
y(t,u), u, \const).\ee
 If we insert $x^{\mu}$ into metric (\ref{metric}), we get the induced
 metric on the world-sheet,
 \be  ds_{induced}^2=\left(
\frac{u}{R_{D4}} \right)^{3/2}\left [
-(\frac{1}{\gamma^{2}}-\frac{u_{T}^{3}}{u^{3}}) dt^2+
(\frac{R_{D4}}{u})^{3}\frac{1-\frac{u^{3}_{T}}{\gamma^{2}u^{3}}du^2}{(1-\frac{u^{3}_{T}}{u^{3}})^{2}}
+\frac{2v^{2}R_{D4}^{3/2}u^{3/2}_{T}dtdu}{u^{3}(1-\frac{u^{3}_{T}}{u^{3}})}+
d\delta y(t,u)^{2}\right] ~, \ee where we have restored the
dimension of the coordinates, and $\gamma$ is the Lorentz factor.
This metric can be simplified by performing coordinate
transformation,
 \be
\begin{aligned}
  \bar{t}&=\frac{t}{\gamma} + \frac{ R_{D4}^{3/2}}{3\gamma
  u^{1/2}_{T}}\left[\ln\frac{p-1}{\sqrt{p^{2}+p+1}}\right.\\
  &\qquad  \left.- \sqrt{3}\arctan\frac{2p+1}{\sqrt{3}}-\gamma^{2/3}\ln
 \frac{q-1}{\sqrt{q^{2}+q+1}} +
 \gamma^{2/3}\sqrt{3}\arctan\frac{2q+1}{\sqrt{3}}\right],
\end{aligned}
 \ee
 \be \bar{u} = u ,\ee
 where $p = \frac{u}{u_{T}}$, $q = \frac{u}{\gamma^{2/3}u_{T}}$, then the metric becomes
 \be \label{inducedmetric} ds_{induced}^2=\left(
\frac{\bar{u}}{R_{D4}} \right)^{3/2}\left [
-(1-\frac{\gamma^{2}u_{T}^{3}}{\bar{u}^{3}}) d\bar{t}^2+
\frac{(\frac{R_{D4}}{\bar{u}})^{3}d\bar{u}^2}{1-\frac{\gamma^{2}u^{3}_{T}}{\bar{u}^{3}}}
+ d\delta y(\bar{t},\bar{u})^{2}\right]  .\ee
 Now we define $\hat{t}\gamma^{2/3}u_{T} = \bar{t}$,~$\hat{u}\gamma^{2/3}u_{T} =\bar{u} $,
 ~$\delta \hat{y}(\hat{t},\hat{u})\gamma^{2/3}u_{T}=\delta y(\bar{t},\bar{u})$,
 $\hat{R}\gamma^{2/3}u_{T} = R_{D4}$, then the metric looks like the original one
 (\ref{metric}),
\be  ds^2=(\gamma^{2/3}u_{T})^{2}(\hat{R}/\hat{u})^{-3/2}\left [
-f(\hat{u}) d\hat{t}^2 +
(\frac{\hat{R}}{\hat{u}})^{3}\frac{d\hat{u}^2}{f(\hat{u})} + d\delta
\hat{y}(\hat{t},\hat{u})^{2} \right ]~, \ee
 with
 $u_{T}\rightarrow \gamma^{2/3}u_{T}$, $R\rightarrow \hat{R}$. A
 compelling characteristic is that from the point of view of string
 world-sheet, the horizon shifts to $\hat{u} = 1$, or $u$=$\gamma^{2/3}$.
 This new horizon is usually called the world-sheet horizon.
 The near horizon Kruskal coordinates can be defined by
 \be \hat{U} = -4\hat{M}e^{-(\hat{t}-\hat{r}_{*})/4\hat{M}},\quad \hat{V} = 4\hat{M}e^{(\hat{t}-\hat{r}_{*})/4\hat{M}}~,\ee
 where \be  \hat{\rho} = \frac{\hat{R}^{3/2}}{3}\hat{u},\quad 2\hat{M} = \frac{\hat{R}^{3/2}}{3}~,\ee
 \be\hat{r}^{*} = \hat{\rho} + 2\hat{M} \ln |\hat{\rho}/2\hat{M} - 1|~.\ee
\begin{figure}[ht]
\begin{center}
\vspace{0ex}
\includegraphics[width=.4\textwidth]{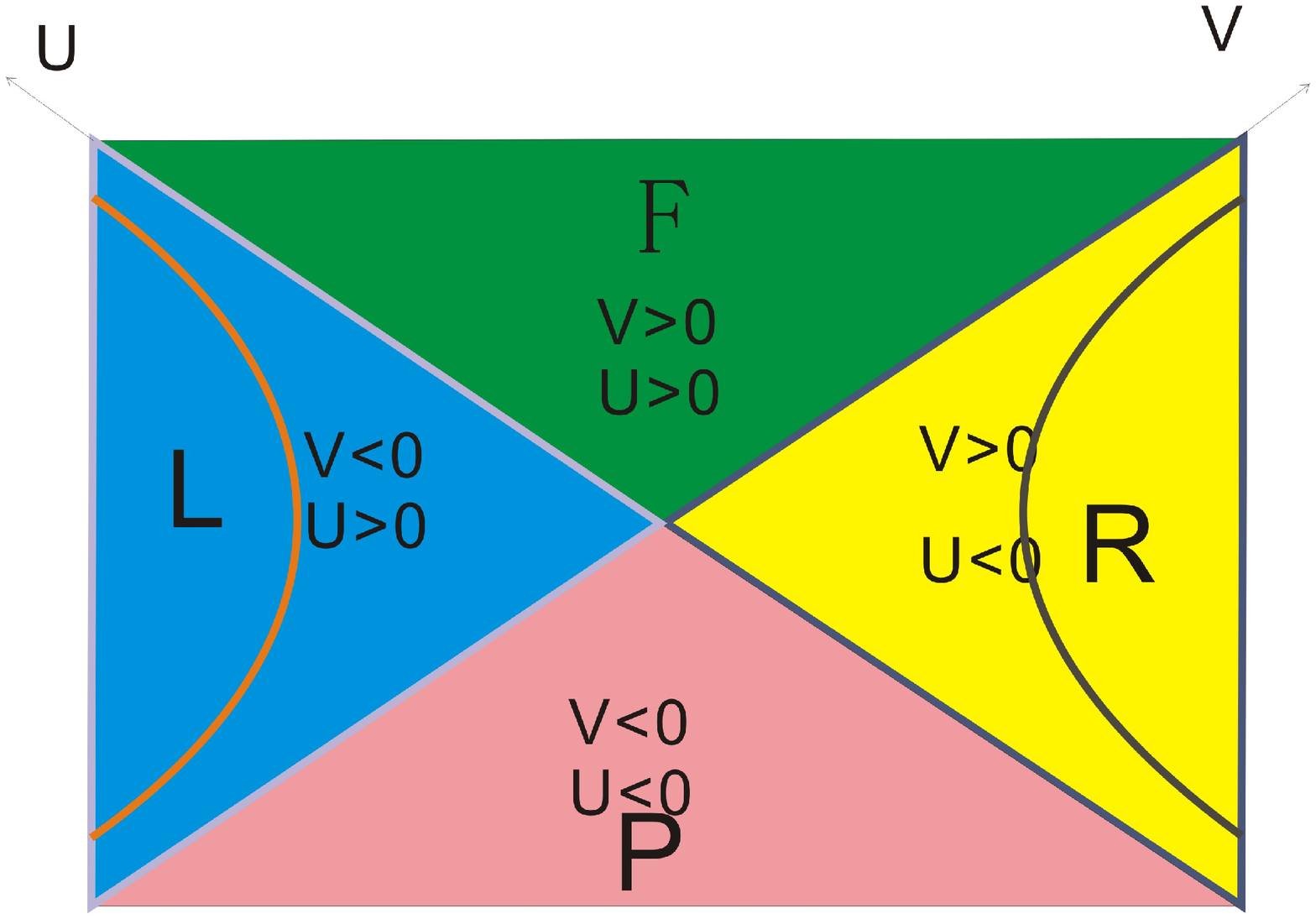}
\includegraphics[width=.32\textwidth]{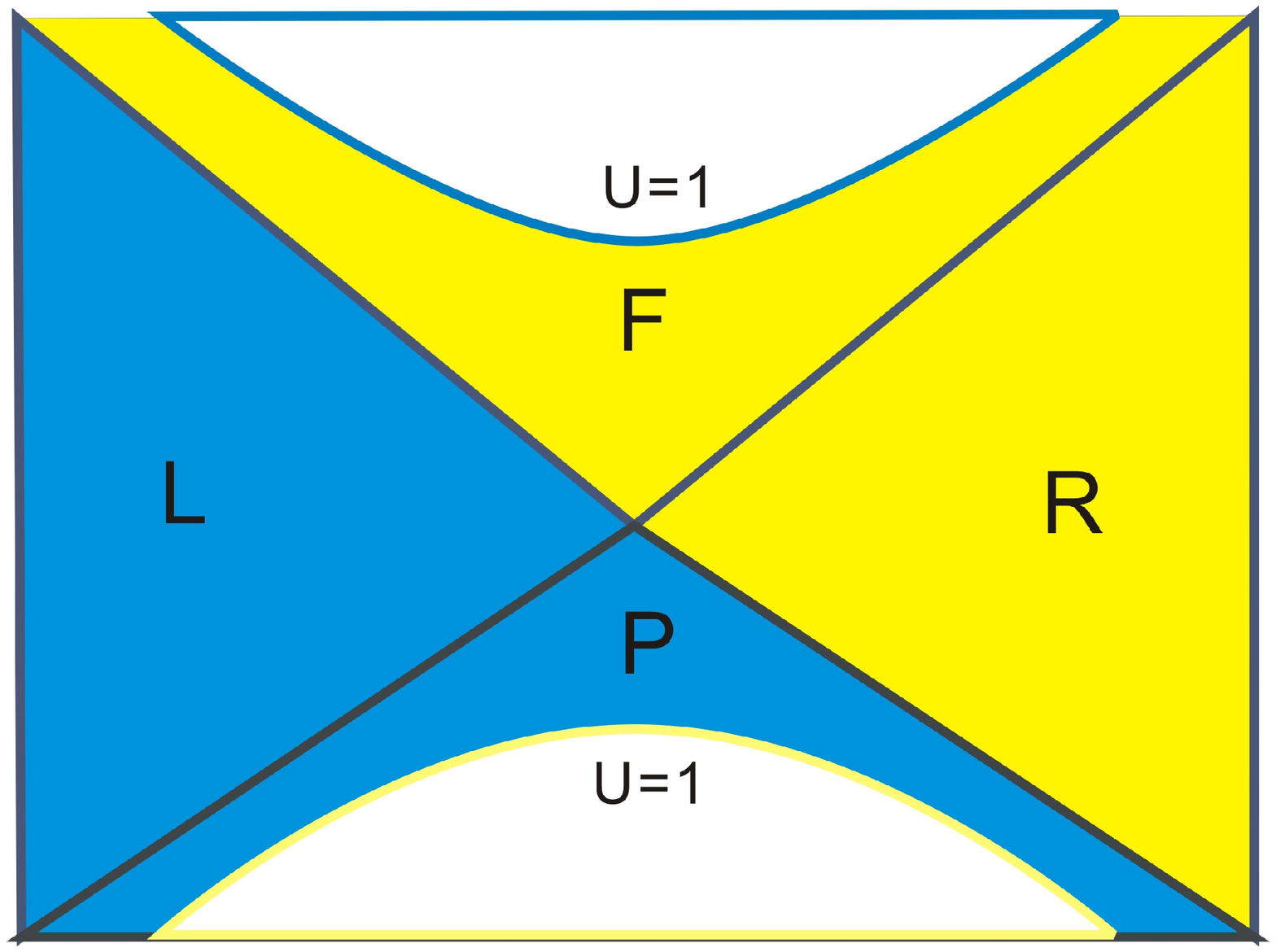}
\end{center}
\caption{The left figure is the space-time Penrose diagram in
Kruskal coordinates. The right figure is the string world-sheet
Penrose diagram in world-sheet Kruskal coordinates. The regions with
yellow color in two figures, represent the same zone in space-time,
so are the blue regions. The hyperbolas in the space-time Penrose
diagram correspond to the world-sheet horizon.}
\end{figure}

 In Figure 2, in the R patch of world-sheet Kruskal plane, the action of the small
 fluctuations $\delta \hat{y}(\hat{t},\hat{u})$ is derived from the
 Nambu-Goto action
 \bea
  S = \frac{(\newhorizon)^{2}}{\tension} \int d\hat{t} d\hat{u} \sqrt{1-f^{-1}(\hat{u})\delta\dot{
  \hat{y}}^{2}+f(\hat{u})\frac{\hat{u}^{3}}{\hat{R}^{3}}\delta \hat{y}'^{2}},
 \eea
 where ``dot'' denotes $\partial_{\hat{t}} $, ``prime'' denotes $\partial_{\hat{u}}$. Because $\delta \hat{y}$ is small, we can expand action around
$\delta
 \hat{y}$ = 0, and keep up to the second order of $\delta \hat{y}$. \be
  S = \frac{(\newhorizon)^{2}}{\tension} \int d\hat{t} d\hat{u} (1-\frac{1}{2}f^{-1}(\hat{u})\delta\dot{
  \hat{y}}^{2}+\frac{1}{2}f(\hat{u})\frac{\hat{u}^{3}}{\hat{R}^{3}}\delta
  \hat{y}'^{2}).
 \ee
 Note that the infinite part of the action is subtracted
 since it appears in the numerator and denominator of
 Eq. (\ref{gauge-string}).
 To solve fluctuation $\delta \hat{y}$ ,
 we define \be \delta \hat{y}(\hat{t},\hat{u}) = \int \frac{d\hat{\omega}}{2\pi}
 \exp^{-i\hat{\omega}\hat{t}}\hat{y}(\hat{\omega})\hat{Y}(\hat{u},\hat{\omega}),\ee
 where we have chosen to normalize $\hat{Y}(\hat{u}=\infty,\hat{\omega})$
 = 1, because $\hat{y}(\hat{\omega})$ is the Fourier
 transformation of the boundary value of $\delta
 \hat{y}(\hat{t},\hat{u})$.
 The Euler-Lagrange equation of small string fluctuations can be
 written as
 \be\label{EOM} \partial^{2}_{\hat{u}}\hat{Y}_{\hat \omega} + \frac{3\hat{u}^{2}}{\hat{u}^{3}-1}\partial_{\hat{u}}\hat{Y}_{\hat
 \omega}+ \frac{\hat{\omega}^{2}\hat{u^{3}}\hat{R}^{3}}{(\hat{u}^{3}-1)^{2}}\hat{Y}_{\hat
 \omega} = 0.\ee
 This equation is solved by
 \be \hat{Y}(\hat{u},\hat{\omega}) =
 (1-\frac{1}{\hat{u}^{3}})^{-i\frac{\hat{\omega}\hat{R}^{3/2}}{3}}F(\hat{\omega},\hat{u}),\ee
 where $F(\hat{\omega},\hat{u})$ is a regular function of $\hat{u}$.
 $(1-\frac{1}{\hat{u}^{3}})^{-i\frac{\hat{\omega}\hat{R}^{3/2}}{3}}$
 corresponds to in-falling fluctuation in the world-sheet horizon $\hat{u}=1$. The complex
 conjugate of this expression is also a solution of the differential
 equation (\ref{EOM}) and corresponds to out-going fluctuation in the horizon.

 Now, we have obtained the solution in the R patch of world-sheet
 Kruskal plane representing the right part of the R patch of
 the space-time Kruskal plane. The right and left parts in the R patch
 are separated by the curve $u=\gamma^{2/3}$, as in Figure 2. But in
order to use gauge/string duality to obtain the generating
functional of the Feynman correlation function, we need to know the
solution defined in the whole Kruskal plane of space-time. To this
goal, we will extend the solution in the right half of the R patch
into other parts of Kruskal plane one by one. Firstly, we should
extend the solution into the whole R patch of space-time Kruskal
plane. In other words, this amounts to extending the solution from R
patch of world-sheet Kruskal plane to the R and F parts of
world-sheet Kruskal plane. In terms of near horizon Kruskal
coordinates, the in-falling and out-going solutions in R patch of
world-sheet Kruskal plane behave as
 \bea\label{in-falling}
 \mbox{in-falling:}~~~~   e^{-i\hat{\omega}\hat{t}}\hat{Y}(\hat{\omega},\hat{u})\sim e^{-4i\hat{\omega}\hat{M}\ln\hat{V}},
 \eea
 \bea\label{out-going}
 \mbox{out-going:}~~~~   e^{-i\hat{\omega}\hat{t}}\hat{Y}^{*}(\hat{\omega},\hat{u})\sim e^{4i\hat{\omega}\hat{M}\ln-\hat{U}}.
 \eea
 Because the Kruskal coordinates are global, actually, in the F
 patch of world-sheet Kruskal plane, near world-sheet horizon, we can also write down
 these two solutions as
\bea\label{in-falling}
 \mbox{in-falling:}~~~~   e^{-i\hat{\omega}\hat{t}}\hat{Y}(\hat{\omega},\hat{u})\sim e^{-4i\hat{\omega}\hat{M}\ln\hat{V}},
 \eea
 \bea\label{out-going}
 \mbox{ out-going:}~~~~   e^{-i\hat{\omega}\hat{t}}\hat{Y}^{*}(\hat{\omega},\hat{u})\sim e^{4i\hat{\omega}\hat{M}\ln\hat{U}}.
 \eea
From above expressions, in the F patch of world-sheet Kruskal plane,
 we see that the in-falling solution is still effective, because
$\hat{V}>0$, but $\hat{U}$ changes sign. So we will do an analytic
extension for the out-going solution to make it an solution in the R
and F patches of the world sheet Kruskal plane. Following Herzog and
Son's prescription \cite{Herzog and Son}, the out-going solution
should cross the horizon from the upper half of complex $\hat{U}$,
this results in that the out-going wave should picks up a factor
$e^{4 \pi \hat{\omega}\hat{M}}$. Physically, this indicates that the
out-going wave should be purely negative-frequency. We can repeat
this process in the P and L patches. Having done this, the out-going
solution picks up a factor $e^{-4 \pi \hat{\omega}\hat{M}}$. Now, we
have solutions defined in $\hat{V}>0$, $\hat{V}<0$ parts of the
world-sheet Kruskal plane. As we have demenstrated previously, the R
and F patches of the world sheet Kruskal plane represent the R patch
of the space-time Kruskal plane. The L and P patches of the world
sheet Kruskal plane represent the L patch of the space-time Kruskal
plane. So we know how these solutions behave in the R and L patches
of the space-time Kruskal plane respectively. Near space-time
horizon, in terms of the space-time near horizon Kruskal
coordinates, in the R patch of the space-time
 \be\label{R-patch1}
  \mbox{in-falling:}~~~~   e^{-i\hat{\omega}\hat{t}}\hat{Y}(\hat{\omega},\hat{u})\sim e^{-4i\omega M\ln V},
 \ee
\be\label{R-patch2}
  \mbox{out-going:}~~~~   e^{4\pi \hat{\omega}\hat{M}}e^{-i\hat{\omega}\hat{t}}\hat{Y}^{*}(\hat{\omega},\hat{u})\sim e^{4\pi \hat{\omega}\hat{M}}e^{-4i\omega M\ln V}.
 \ee
 in the L patch of the space-time
\be\label{L-patch1}
  \mbox{in-falling:}~~~~   e^{-i\hat{\omega}\hat{t}}\hat{Y}(\hat{\omega},\hat{u})\sim e^{-4i\omega M\ln(-V)},
 \ee
\be\label{L-patch2}
  \mbox{out-going:}~~~~   e^{-4\pi \hat{\omega}\hat{M}}e^{-i\hat{\omega}\hat{t}}\hat{Y}^{*}(\hat{\omega},\hat{u})\sim e^{-4\pi \hat{\omega}\hat{M}}e^{-4i\omega M\ln(-V)}.
 \ee
 These expressions tell us that in the point of view of space-time,
 the in-falling and out-going solutions in the world-sheet are both
 in-falling waves. Moreover, the R patch solutions can be
 interpreted as the solution in both the R and F patches. The L
 patch solutions can be interpreted as the solution in both the L
 and P patches, because $V>0$ in both the R and F patches; $V<0$ in both the L
 and P patches. So far, we have obtained the solution defined
 in the whole $V>0$ part of Kruskal plane, and the solution defined in
 the whole $V<0$ part of Kruskal plane. They can be deduced from the
 following four different solutions defined in R and L patches.
 \be\label{IN}
\y_{ R,~i} = \left\{ \begin{array}{ll}
e^{-i\w \t}\Y(\w,\u) & \mbox{in R} \\
0 & \mbox{in L}
\end{array}
\right. \; \; \; \; \; \; \; \; \; \;
\y_{ L ,~i} = \left\{ \begin{array}{ll}
0 & \mbox{in R} \\
 e^{-i\w \t}\Y(\w,\u)& \mbox{in L}
\end{array}
\right. \ , \ee
 \be\label{OUT} \y_{ R ,~o} = \left\{ \begin{array}{ll}
e^{-i\w \t}\Y^*(\w,\u) & \mbox{in R} \\
0 & \mbox{in L}
\end{array}
\right.   \; \; \; \; \; \; \; \; \; \;
\y_{ L ,~o} = \left\{ \begin{array}{ll}
0 & \mbox{in R} \\
 e^{-i\w \t}\Y^*(\w,\u)& \mbox{in L}
\end{array}
\right. \, . \ee
 Following the Herzog and Son prescription \cite{Herzog and Son}, we
 look for linear combinations of these expressions that, close to
 the horizon, are analytic in the lower half of the complex V
 plane. Physically, this means that the
in-falling wave should be purely positive-frequency. With this
requiriement, the two linearly independent combination are:
\be\label{combine} \y_o=\y_{R,~o}+ \alpha_o \y_{L,~o} \, ,\\~~~~
\y_i=\y_{R,~i}+ \alpha_i \y_{L,~i} \, . \ee where the $\alpha_o$ and
$\alpha_i$ can be determined from the near horizon behaviors of
these
 solutions Eqs. (\ref{R-patch1}), (\ref{R-patch2}), (\ref{L-patch1}) and (\ref{L-patch1})
 \be\label{a1}\alpha_o= e^{8\pi \w\M}e^{-4\pi\omega M},\ee
 \be\label{a2} \alpha_i= e^{-4\pi\omega M}.\ee
 These two solutions are used as basis for the linearized
string fluctuations defined over the full (AdS) Kruskal plane
\be\label{solution} \y(\t,\u)=\int \frac{d\w}{2\pi}
\left(a(\omega)\y_o(\omega)+ b(\omega)\y_i(\omega)\right)\, . \ee
The coefficients $a(\w),~b(\w)$ can be determined by the boundary
values of the solutions. Because we have \bea
\y(\t,\u=\infty)\big|_{R}&=&\int \frac{d\w}{2\pi}e^{-i\w \t} \y_1(\w) \, ,\\
\y(\t,\u=\infty)\big|_{L}&=&\int \frac{d\w}{2\pi}e^{-i\w \t}
\y_2(\w) \, . \eea We obtain
  \be\label{aw} a(\w)=\hat{n}(-\y_{1}(\w)+e^{4\pi\omega M}\y_{2}(\w)),\ee
  \be \label{bw} b(\w)=\hat{n}(e^{8\pi\w\M}\y_{1}(\w)-e^{4\pi\omega M}\y_{2}(\w)).\ee
 where $\hat{n} = 1 /(e^{8\pi\w\M}-1)$.
 Now we compute the boundary action in terms of the string
 solution: In $(\t,\u)$ coordinates
 \bea
 S_B=\frac{(\newhorizon)^2}{\tension}
   \left[
   \int_R \frac{d\hat{\omega}}{2\pi} f(\u)\frac{\u^{3}}{\hat{R}^{3}}
        \y(-\hat{\omega},\hat{u})\partial_{\hat{u}} \y(\hat{\omega},\hat{u})
   -
   \int_L \frac{d\hat{\omega}}{2\pi} f(\u)\frac{\u^{3}}{\hat{R}^{3}}
        \y(-\hat{\omega},\hat{u})\partial_{\hat{u}} \y(\hat{\omega},\hat{u})
   \right]  .\,\eea
Notice that $\w=\gamma\omega$, $\y_{1}(\w)=
\gamma^{-1}y_{1}(\omega)$, $\y_{2}(\w)= \gamma^{-1}y_{2}(\omega)$,
and using Eqs. (\ref{IN}), (\ref{OUT}), (\ref{a1}), (\ref{a2}),
(\ref{aw}), (\ref{bw}) and (\ref{solution}), this action can be
expressed as
 \be
 \begin{aligned}
&S_B=\frac{(\newhorizon)^2}{ \tension\gamma} \int
\frac{d\omega}{2\pi} \left.\Big[ \right.\\
& y_1(-\omega)y_1(\omega)\left.(
(\n+1)f(\u)\frac{\u^{3}}{\hat{R}^{3}}
\Y^*(-\hat{\omega},\hat{u})\partial_{\hat{u}}
\Y(\hat{\omega},\hat{u})
 -\n f(\u)\frac{\u^{3}}{\hat{R}^{3}}
 \Y(-\hat{\omega},\hat{u})\partial_{\hat{u}} \Y^*(\hat{\omega},\hat{u})
                         \right)\\
& +y_1(-\omega)y_2(\omega)e^{\pi\omega/2}\n
      \left.( -f(\u)\frac{\u^{3}}{\hat{R}^{3}}
   \Y^*(-\hat{\omega},\hat{u})\partial_{\hat{u}} \Y(\hat{\omega},\hat{u})
+f(\u)\frac{\u^{3}}{\hat{R}^{3}}
   \Y(-\hat{\omega},\hat{u})\partial_{\hat{u}} \Y^*(\hat{\omega},\hat{u})
      \right)\\
& +y_2(-\omega)y_1(\omega)e^{-\pi\omega/2}(1+\n)
      \left.( -f(\u)\frac{\u^{3}}{\hat{R}^{3}}
   \Y^*(-\hat{\omega},\hat{u})\partial_{\hat{u}} \Y(\hat{\omega},\hat{u})
+f(\u)\frac{\u^{3}}{\hat{R}^{3}}
   \Y(-\hat{\omega},\hat{u})\partial_{\hat{u}} \Y^*(\hat{\omega},\hat{u})
      \right)\\
&+ y_2(-\omega)y_2(\omega)\left.( \n f(\u)\frac{\u^{3}}{\hat{R}^{3}}
 \Y^*(-\hat{\omega},\hat{u})\partial_{\hat{u}} \Y(\hat{\omega},\hat{u})
 -(\n+1)f(\u)\frac{\u^{3}}{\hat{R}^{3}}
 \Y(-\hat{\omega},\hat{u})\partial_{\hat{u}} \Y^*(\hat{\omega},\hat{u})
                        \right)\Big]\big|_{u\rightarrow +\infty}  .
\end{aligned}
\ee
 From this expression we can read off the Feynman correlation by
 taking derivatives with respect to $y_1(-\omega)$, $y_1(\omega)$.
 \be
  G_{F}(\omega)=\frac{1}{\tension \gamma
  \newhorizon}[{(\n+1)f(\u)\frac{\u^{3}}{\R^{3}}\Y^{*}\partial_{\u}\Y(\w,\u)-\n
  f(\u)\frac{\u^{3}}{\R^{3}}\Y\partial_{\u}\Y^{*}(\w,\u)}]\big|_{\u\rightarrow\infty},
  \ee
 where we have restored physical dimension of $G_{F}$. When $\w\rightarrow 0$ we can expand $\Y(\w,\u)$ in a power series
 in $\w$ and solve order by order
  \be
  \begin{aligned}
  \Y(\w,\u) &=(1-\frac{1}{\u^{3}})^{-2i\M\w}\left[ 1-2i\M\w(3\ln(\u/\sqrt{\u^{2}+\u+1})\right. \\
  &\qquad \left.-\sqrt{3}\arctan(2\u/\sqrt{3}+1/\sqrt{3})+\frac{\sqrt{3}\pi}{2})+{\cal O}(\w^{2})\right].
\end{aligned}
  \ee
 So
  \be
  \kappa_T=-\lim_{\omega \to 0}\Im G_F(\omega)=\frac{3\gamma^{1/3}u_{T}^{2}}{4\pi^{2}\alpha^{'}R_{D4}^{3}},
\ee
 where we have used $\R^{3}=\gamma^{-2}R^{3}$.
 Using the parameter relation between string  theory and gauge
 theory, we obtain
  \be \kappa_T=\frac{16\sqrt{2}\pi}{27}\frac{\gamma^{1/3}\lambda
  T^{4}}{T_{d}},
  \ee
  where $\lambda$ is the the 't~Hooft coupling of YM, $g_{YM}^2
  N_{c}$.
 When $v\rightarrow 0$ if $\kappa_T$ and the drag coefficient which is often
denoted as $\eta_{D}$ satisfy the Einstein relation, then the drag
coefficient in S-S model should be
 \be \eta_{D}=\frac{1}{\tension M}(\frac{u_{T}}{R_{D4}})^{3/2}.\ee
This is the result appearing in \cite{drag-Talavera}.

\section{An estimation of $\hat{q}$ with respect to gluon energy loss in S-S model}
\subsection{The initial energy-momentum of gluon in S-S model}
In \cite{Gubser2}, S.~Gubser and his collaborators proposed a simple
but interesting idea to estimate the jet quenching parameter $\q$,
by computing how far an off-shell gluon propagates in the finite
temperature $\mathcal{N}=4$ SYM before it loses all energy and
resolves into the medium. Their estimation is mainly based on an
extension of BDMPS formalism \cite{BDMPS,Zakharov1,Zakharov2} from
light-like parton to time-like parton, which is to replace the light
cone distance $L^-$, by $\sqrt{2}\Delta x$, where $\Delta x$ is the
parton's in-medium space distance, often called penetration length.
In short, this is
 \be \Delta E=\frac{1}{4}\alpha_s C_R \q \frac{L^{-2}}{2}\rightarrow\Delta E=\frac{1}{4}\alpha_s C_R \q \Delta
 x^{2},\ee
 and
 \be \q= \frac{4\Delta E}{\alpha_s C_R \Delta x^2}.
 \ee
 In above expressions, $\alpha_s$ is the strong coupling constant, and $C_R$ is the
 color group $SU(N)$ Casimir $C_{2}(R)$ evaluated in the parton's
 representation, for gluon $C_R$=N.
Although this is not an exact calculation about $\q$, it has a
striking virtue appealing to us, that is their result $\q\sim21{\rm
GeV^{2}}/{\rm fm}$ lies within the $3\sigma$ range of averaged $\q$
\cite{average-q}, while other people's result \cite{jetquench-LRW1}
fails. The $3\sigma$ range of averaged $\q$ is
 \be
 7 \frac{\rm GeV^{2}}{\rm fm} \lesssim\langle \hat{q} \rangle \lesssim
    28 \frac{\rm GeV^2}{\rm fm},
    \ee
with lowest $\chi^2$ at $\langle \hat{q} \rangle \approx 13\,{\rm
GeV}^2/{\rm fm}$. S-S model is also a holographic dual to QCD, so we
can ask whether their method still has this advantage in S-S model.

To answer this question, we do the following estimate in S-S model.
Firstly, we will introduce the following background for the
simplicity of calculation,
 \be \label{newbackground}ds^{2}=u^{2}_{T}(R y)^{-\frac{3}{2}}[-(1-y^{3})dt^{2} + d\vec{x}^{2}+\frac{R^{3}dy^{2}}{y(1-y^{3})}].
 \ee
This is obtained from the metric (\ref{metric}), by replacing $u$
with $\frac{1}{y}$ and letting $x_{4},~\Omega_{4}$ be constant.
Following S.~Gubser's approach, a gluon is represented by a doubled
string which rises from the horizon up to a minimum $ y_{\rm UV}$,
and then falls back down to the horizon as in Figure 3.
\begin{figure}[ht]
\begin{center}
\vspace{1ex}
\includegraphics[width=.43\textwidth]{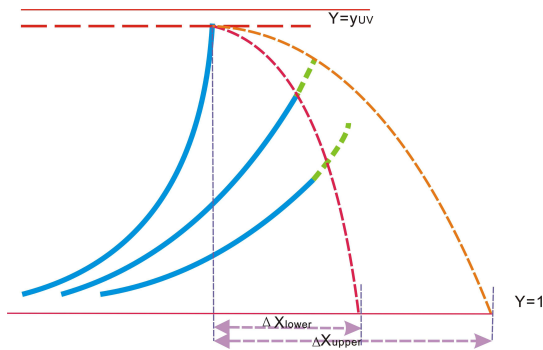}
\end{center}
\caption{The blue line shows the shape of a doubled string which
represents a gluon. The pink line and the orange line represent the
world-sheet light signal's trajectory and the massless particle's
trajectory we will introduce respectively. We will use these two
trajectories to give an estimation of the range of gluon penetration
length. }
\end{figure}

Required to be stable, the gluon's initial state is constructed from
the trailing string solution, the trailing string is a string moving
with constant velocity in $x^{1}$ direction in the background
(\ref{newbackground}), and its shape is
 \bea\label{trailing2}
 x^{1}(t)&=&vt+
   vR^{\frac{3}{2}}\left[\frac{\ln(1-y)}{(1-\alpha)(1-\beta)}+\frac{\ln(1-\beta y)}{(\alpha-1)(\alpha-\beta)}+\frac{\ln(1-\alpha y)}{(\beta-\alpha)(\beta-1)}\right]+\nonumber
\\
&&  vR^{\frac{3}{2}}\left[\frac{\ln(1-\beta
y)}{\alpha-\beta}-\frac{\ln(1-\alpha y)}{\alpha-\beta}\right] .\eea
 This solution is equal to (\ref{trailing}), $\alpha=e^{2\pi i/3}$
 and $\beta$ is $\alpha$'s complex conjugate, we prefer this
 expression because of its convenience for following calculation.
 So, if we insist to use world-sheet coordinates  $\sigma^\alpha = (t,y)$, initially, the induced metric on the world-sheet is
 \be
 g_{\alpha\beta} =u_{T}^2 (R y)^{-\frac{3}{2}} \begin{pmatrix}
    -h+v^2 & -v^2 yR^{3/2}/h \\ -v^2 yR^{3/2}/h &
     R^{3}(h+v^2-hv^2)/h^2y \end{pmatrix},
     \ee
where $h=1-y^{3}$. The world-sheet current density of
energy-momentum

 is
  \be
  P^\alpha_m = {1 \over \pi\alpha' h (1-v^2)}
   \begin{pmatrix} -h-v^2+hv^2 & v & 0 & 0  \\
    -h v^2 y^2R^{-3/2} & hv y^2R^{-3/2} & 0 & 0  \end{pmatrix} ,\
\ee
 where m=(0,~1,~2,~3) corresponds to $(t,\vec{ x})$. Usually, the
 related doubled string's energy-momentum in these four dimensions is
 \be \label{energy-momentum1}p_m = \frac{1}{u_T}\int_{y_{\rm UV}}^1 dy \, \sqrt{-g} P^t_m=
   {u_T \over \pi\alpha' } {1 \over \sqrt{1-v^2}} \int_{y_{\rm UV}}^1
    {dy \over hy^2} \begin{pmatrix} -h-v^2+hv^2 & v & 0 & 0  \end{pmatrix} .\
\ee Because $\partial_{t}$, $\partial_{\vec{x}}$ is the Killing
vectors, so $p_{m} $ can be identified with the four-momentum of the
gluon in the boundary gauge theory. But we are also confronted with
the problem appearing in $\mathcal{N}$=4 SYM: the energy-momentum
has a logarithmic divergence at $y$=1. Gubser gives an explanation
for the appearance of this kind of divergence: this divergence is
due to the fact that, to form the shape of a trailing string needs
infinitely long time, and during this period, infinite
energy-momentum has been transferred from the string to the medium,
but it is still contained in the right hand side of Eq.
(\ref{energy-momentum1}). So once this shape has been formed, the
rest energy-momentum of the string should not include the
energy-momentum transferred into the horizon. To make this
subtraction, we compute
 \be \label{energy-momentum2}
\begin{aligned}
 p^{\rm fixed x^{1}}_m &= \frac{1}{u_T}\left.\int_{y_{\rm UV}}^1 dy \, \sqrt{-g}[P^t_m-(\frac{\partial t}{\partial
 y})_{x_1}P^y_m\right.]\\
 & \left.= \frac{u_{T}}{\pi\alpha'}(\frac{1}{y_{UV}}-1)(-1, v, 0,
 0),\right.
\end{aligned}
\ee where the subscript $x^{1}$ indicates that the above integral is
carried out along the $x^{1}$=conctant contour in the ($t$,$y$)
plane. With Green theorem and $P^{\alpha}_{m;~\alpha}$=0, it is not
hard to prove the difference between (\ref{energy-momentum1}) and
(\ref{energy-momentum2}) is the amount of energy-momentum we want to
subtract from (\ref{energy-momentum1}). Now, we can show to the
reader that this gluon is a time-like one. This is obvious, since
 \be
 E^{2}-\vec{P}^{2}=[\frac{u_{T}}{\pi\alpha'}(\frac{1}{y_{UV}}-1)]^{2}>0.
 \ee
\subsection{Estimation of gluon penetration length}
Since this doubled string's tip does not attach to the boundary
brane, this string will fall down toward horizon. At some moment,
the tip will touch the horizon, from the beginning to this moment,
the tip travels $\Delta x$ in $x^{1}$ direction. Generally, $\Delta
x$ is function of $ y_{\rm UV}$ and velocity $v$ or $\gamma$. For a
fixed energy gluon, $\Delta x$ is function of $ y_{\rm UV}$ or $v$,
for example, we choose $\Delta x$=$\Delta x(v)$, then the maximum
value of $\Delta x(v)$ with respect to $v$ can be interpreted as the
penetration length of the gluon. Although, we know the initial shape
of the doubled string, it is difficult to compute $\Delta x$ in
terms of the EOM of string, not to mention $\Delta x_{\rm max}$. But
with the methods proposed by Gubser, we can find out a lower and an
upper bound for  $\Delta x_{\rm max}$.

Firstly, we consider the lower bound. The initial shape of the
doubled string can be interpreted as cutting the trailing string
which attaches to the boundary brane, at $y=y_{\rm UV}$ at some
time. Meanwhile, a light signal is emitted from the cut. Since the
disturbance arising from cutting the string cannot propagate more
quickly than light, the string will keep its shape where the light
 has not even arrived, as if we did not make such a cut.
The displacement of light in $x^{1}$ direction, denoted as $\Delta
x_{\rm low}$, should be the lower bound of $\Delta x$, then $\Delta
x_{\rm low,~max}$ serves as the lower bound of $\Delta x_{\rm max}$.
The trajectory of the light signal can be determined from the
light-like tangent vector of world-sheet metric, which satisfies as
$t$ increases, $y$ increases. There is only one light-like tangent
vector field meeting this requirement. It is \be\label{light-signal}
l^\alpha =
\begin{pmatrix}
R^{\frac{3}{2}}[-\frac{v^{2}y}{h(h-v^{2})}+\frac{\sqrt{1-v^{2}}}{\sqrt{y}(h-v^{2})} ]\\[3pt] 1 \end{pmatrix}
.\ee So the light signal's trajectory is determined by
 \be \frac{dt}{dy}=R^{\frac{3}{2}}[-\frac{v^{2}y}{h(h-v^{2})}+\frac{\sqrt{1-v^{2}}}{\sqrt{y}(h-v^{2})}]. \ee
 Solving this differential equation, we obtain
 \bea\label{light-trace}
&&t=R^{\frac{3}{2}}\gamma^{\frac{2}{3}}\left[\frac{2}{(1-\alpha)(1-\beta)}\ln(1+\gamma^{1/3}\sqrt{y})+\frac{1}{(\alpha-1)(\alpha-\beta)}\ln\frac{1+\beta\gamma^{1/3}\sqrt{y}}{1-\beta\gamma^{1/3}\sqrt{y}}\right]\nonumber
\\
&&+R^{\frac{3}{2}}\gamma^{\frac{2}{3}}\left(\frac{1}{(\beta-\alpha)(\beta-1)}\ln\frac{1+\alpha\gamma^{1/3}\sqrt{y}}{1-\alpha\gamma^{1/3}\sqrt{y}}\right)\nonumber
\\ &&
-R^{\frac{3}{2}}\left[\frac{\ln(1-y)}{(1-\alpha)(1-\beta)}+\frac{\ln(1-\beta
y)}{(\alpha-1)(\alpha-\beta)}+\frac{\ln(1-\alpha
y)}{(\beta-\alpha)(\beta-1)}\right]\nonumber
\\
&&  -R^{\frac{3}{2}}\left[\frac{\ln(1-\beta
y)}{\alpha-\beta}-\frac{\ln(1-\alpha
y)}{\alpha-\beta}\right]\nonumber
 \\&& R^{\frac{3}{2}}\gamma^{\frac{2}{3}}\left[\frac{\ln(1-\gamma^{2/3}\beta
y)}{(\alpha-1)(\alpha-\beta)}+\frac{\ln(1-\gamma^{2/3}\alpha
y)}{(\beta-\alpha)(\beta-1)}\right]\nonumber
\\
&&
+R^{\frac{3}{2}}\gamma^{\frac{2}{3}}\left[\frac{\ln(1-\gamma^{2/3}\beta
y)}{\alpha-\beta}-\frac{\ln(1-\gamma^{2/3}\alpha
y)}{\alpha-\beta}\right] ,\eea
 where the $\alpha$ and $\beta$ are the same as Eq. (\ref{trailing2}). Plugging Eq. (\ref{light-trace}) into Eq. (\ref{trailing2}) we obtain the orbit
\be
   x^{1}(z)=u_{T}\frac{2}{3}vR^{\frac{3}{2}}\gamma^{\frac{2}{3}}[\ln(1+\gamma^{1/3}\sqrt{y})+\alpha\ln(1+\beta\gamma^{1/3}\sqrt{y})+\beta\ln(1+\alpha\gamma^{1/3}\sqrt{y})].
\ee
 So \be \Delta  x _{\rm low}=  x^{1}(1)- x^{1}( y_{ UV}),\ee
 where we restore the physical dimension of $x^{1}$. For convenience, we define
 \be \Delta\x_{\rm low}=\frac{\Delta  x
 _{\rm low}}{u_{T}R^{\frac{3}{2}}},\ee
 \be \E=\frac{\pi\alpha' E}{u_{T}}.\ee
 Now we should find the maximum value of $\Delta\x_{\rm low}$ with a
 fixed $\E=\gamma(\frac{1}{y_{\rm UV}}-1)$.
 If we define
 \be \xi=\gamma^{2/3} y_{UV},\ee
 then $\Delta\x_{\rm low}$ is a function of $\xi$.
Usually, to find the maximum value of $\Delta\x(\xi)_{\rm low}$, we
will first find a $\xi_{*}$ satisfying\be\label{maximum}
\partial_{\xi}\Delta \x(\xi)_{\rm low,~\xi=\xi_{*}}=0.\ee
 But this is only the point making $\Delta\x(\xi)_{\rm low}$ a local maximum or
minimum and may not be the global maximum or minimum. In fact, there
is only one $\xi_{*}$ satisfying Eq. (\ref{maximum}), and
$\Delta\x(\xi_{*})_{\rm low}$ is the global maximum, this is
supported by numerical result. When $\E\gg1$, we find that $\xi_{*}$
can be expanded in terms of $\E^{\frac{1}{5}}$, it is
 \be
 \xi_{*}=0.38036+0.207807\E^{-\frac{2}{5}}+0.07596\E^{-\frac{4}{5}}-0.781227\E^{-1}+{\cal O}(\hat
 E^{-6/5}).\ee
 And
 \be \Delta\x_{\rm low,~max}=0.84423\E^{\frac{2}{5}}-0.80997+0.04178\E^{-\frac{2}{5}}+0.71431\E^{-\frac{3}{5}}+{\cal
 O}(\E^{-\frac{4}{5}})
 .\ee
 We exhibit the comparison of analytic result and numeric result in
 Figure (4). We find that when $\E\gg1$, the analytic result indeed matches the
 numeric result well.

\begin{figure}[ht]
\begin{center}
\vspace{3ex}
\includegraphics[width=.70\textwidth]{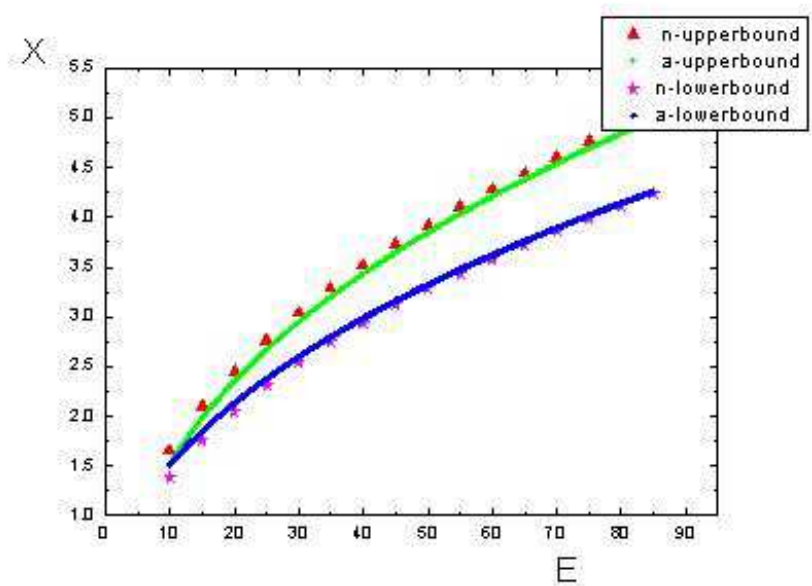}

\end{center}
\caption{The triangles represent the numeric result of upper bound
of penetration length, the stars represent the numeric result of the
lower bound of penetration length, The two lines are the analytic
results of the upper and lower bound of the penetration length for
$\E\gg1$ and their extrapolations to $\E$ not satisfying $\E\gg1$.}
\end{figure}

 Having found the lower bound of penetration length of the
 gluon, next, we shall look for the upper bound. To
 find the upper bound, we use the following picture. When the tip of the
 doubled string begins to fall toward the horizon, it happens that a
 light-like particle is projected from the tip with 5-dimension velocity proportional to the
 5-dimension velocity of the tip. In other words, the massless
 particle's trajectory is tangent to the tip's trajectory  at this point. The
 massless particle moves along the geodesic and will fall farther
 in $x^{1}$ direction than the tip, because the tip is also pulled
 by the rest of the string besides the gravity. So we can perceive
 the massless particle's displacement in $x^{1}$ before it falls into the horizon, as the upper-bound of $\Delta x$ which we
 define before, here we denote this upper-bound by $\Delta x_{\rm upper}$. Then the maximum value of
 $\Delta x_{\rm upper}$ should be the upper-bound of
 $\Delta x_{\rm max}$, for fixed energy. To obtain $\Delta x_{\rm upper}$,
 we solve the EOM of the massless particle in the black hole
 background (\ref{newbackground}) with the initial condition required before. As we know, a massless particle
 can be described by the following action:
 \be S=\frac{1}{2}\int d\eta
 [\frac{1}{e}G_{\mu\nu}\frac{dX^{\mu}}{d\eta}\frac{dX^{\nu}}{d\eta}],\ee
 where $e$ is a Lagrange multiplier, $G_{\mu\nu}$ is
 metric (\ref{newbackground}). In the
 following, let's work in a gauge where $\eta=y$ and consider the
 trajectory of the form
 \be X^{0}=X^{0}(y)~~~X^{1}=X^{1}(y)~~~X^{2}=X^{3}=0.\ee
 Then
 \be S=\int dy \mathcal {L} ~~~\mathcal
 {L}=\frac{u^{2}_{T}(Ry)^{-3/2}}{2e}(-h(X^{0'})^{2}+(X^{1'})^{2}+\frac{R^{3}}{yh}),
 \ee
 where prime denotes $d/dy$. Since the Lagrangian does not contain
 $X^{0}$, $X^{1}$ explicitly, we immediately form two conserved momenta
 \be
 p_{0}=-\frac{u_{T}h(Ry)^{-3/2}X^{0'}}{e}~~~~p_{1}=\frac{u_{T}(Ry)^{-3/2}X^{1'}}{e}.\ee
 The equation of motion of $e$ is a constraint:
 \be\label{constraint} e=\pm\frac{u_{T}}{y^{2}\sqrt{p^{2}_{0}-hp^{2}_{1}}}.\ee
Because of our metric signature, the -$p_{0}$ is energy, and should
be positive, so $p_{0}$ is negative, for $X^{0'}$ is positive,
corresponding that the massless particle falls toward black hole,
finally, we choose plus sign in (\ref{constraint}) for the
trajectory. Then the shape of the trajectories is determined by
\be
\frac{dX^{1}}{dy}=X^{1'}=\frac{ep_{1}(Ry)^{3/2}}{u_{T}}=-R^{3/2}\frac{p_{1}/p_{0}}{\sqrt{y}\sqrt{1-hp^{2}_{1}/p^{2}_{0}}}~,\ee
 where $p_{1}/p_{0}$ is due to the initial condition, since $p_{1}$ and $p_{0}$ are conserved quantities. Because the tip of an open string or a
doubled string must move at the speed of light, at the moment when
the tip is formed, its 5-dimensional velocity should be proportional
to $l^{\mu}|_{y=y_{\rm UV}}$, which is $l^{\alpha}\partial
X^{\mu}/\partial \sigma^{\alpha}|_{y=y_{\rm UV}}$, satisfying
$G_{\mu\nu}l^{\mu}l^{\nu}=0$. For $l^{\alpha}$'s definition, refer
 to (\ref{light-signal}). Then we can derive $p_{1}/p_{0}$
 \be
 p_{1}/p_{0}=v\frac{\sqrt{1-v^{2}}-y^{3/2}_{UV}}{v^{2}y^{3/2}_{UV}-\sqrt{1-v^{2}}(1-y_{UV}^{3})}.\ee
 Now we calculate how far the massless particle propagates in the
 $X^{1}$ direction before falling into the horizon:
 \be \Delta x_{\rm upper}=-R^{3/2}u_{T}\int^{1}_{y_{\rm UV}}dy\frac{p_{1}/p_{0}}{\sqrt{y}\sqrt{1-hp^{2}_{1}/p^{2}_{0}}}.\ee
 The maximum of $\Delta\x_{\rm upper}=\frac{\Delta x_{\rm upper}}{R^{3/2}u_{T}}$ for fixed $\E$ is depicted in the Figure 4. For $\E\gg1$, using the same method as before, we find that
 \be \xi_{*}=0.32141+0.13949\E^{-\frac{2}{5}}+0.03882\E^{-\frac{4}{5}}+{\cal
 O}(\E^{-\frac{6}{5}})
\ee \be\Delta\x_{\rm
upper,max}=0.99033\E^{\frac{2}{5}}-0.79951+0.49157\E^{-\frac{2}{5}}+{\cal
 O}(\E^{-\frac{4}{5}})
.\ee At this moment, we should remind the reader that the analytic
results do not match the numeric results as well as in the
lower-bound case, the deviation may come from the approximate method
we adopt to find out $\xi_{*}$.

So far, we have got two bounds of the gluon penetration length.
Using these two bounds of penetration length to make a rough
estimation of jet quenching parameter will be carried out in the
discussion section.

\section{Discussion}
Following Gubser's description, one is able to extract the jet
quenching parameter from momentum diffusion constant $\kappa_{T}$,
by\be \q_{B}=\frac{2\kappa_{T}}{v},\ee where ``B'' indicates
Brownian motion, because we prefer to interpret it as part of the
jet quenching parameter, this part is due to the Brownian motion
effect. Since strongly coupled S-S model gauge field theory is
different from QCD, there exists considerable uncertainty in how to
translate the results calculated in S-S model into quantitative
predictions in QCD.

To characterize this uncertainty, we recall that in
 $\mathcal{N}$=4 SYM, the optimum scheme is
 \be T_{\mathcal{N}=4}=T_{QCD}/3^{1/4}=280/3^{1/4}~~~~g^{2}_{YM}N_c=5.5~.\ee
The factor $3^{1/4}$ comes from the requirement that $\mathcal{N}$=4
SYM and QCD are compared at the same energy density.
 Similarly, we also require that the S-S model gauge theory and QCD are compared at the same energy density, and then
 choose the following scheme for S-S model:
 \be\label{scheme}
 T_{S-S}=T_{QCD}/\zeta=280{\rm MeV}/\zeta~~~T_{dQCD}=170{\rm MeV}~~~g^{2}_{YM}N_c=5.5~.\ee
In above expressions,
 \begin{equation}\label{zeta}
    \zeta=0.914\lambda^{1/6}(T_{QCD}/T_{d})^{1/3},
 \end{equation}
 and in scheme (\ref{scheme}), $\zeta\sim 1.43$. The explicit calculation of the parameter $\zeta$ will be given in appendix.
 We see that $\zeta$ is an increasing function of $T_{QCD}$. This is
 in accordance with our previous argument that as temperature
 increase, more K-K modes will become active, for their masses are
 in tower of $mT_{d}$, with $m=1,2\cdots$. Since more degrees of
 freedom contribute to the energy density, $\zeta$ must increase
 accordingly to keep the energy density of S-S model plasma equal
 with that of QCD plasma.

 At this moment, we can calculate $\q_{B}$ as following,
 \be
 \q_{B}\sim5.4{\rm GeV^{2}}/{\rm fm}\frac{\gamma^{1/3}}{\zeta^{3}v}=1.85{\rm GeV^{2}}/{\rm fm}\frac{\gamma^{1/3}}{v}.\ee
 For charm quark, $m_{c}$=1.4GeV, typical $p_{c}$=10GeV/c
 \be \label{q}\q_{B}\sim3.5{\rm GeV}^{2}/{\rm fm}.\ee
 This value of $\hat{q}_{B}$ is larger than Gubser's result \cite{Gubser1} $\q_{B}=1.4{\rm GeV}^{2}/{\rm fm}$
 and Liu's result \cite{jetquench-LRW2} $\q=0.86{\rm GeV}^{2}/{\rm
 fm}$.

 Now let us use the gluon penetration length to estimate the
 jet quenching parameter $\q$. Following (\ref{scheme}),
 using BDMPS formalism and setting $\alpha_{s}=1/2$, we find
 \be E=1.115\E {\rm GeV}~~~~\Delta x_{\rm max}(\E)=0.236\Delta\x(\E)_{\rm max} {\rm fm}\ee
 \be \q=\frac{10.350\Delta\E}{\Delta\x(\E)_{\rm max}}{\rm GeV}^{2}/{\rm fm}  .\ee
 Because $\Delta\x(\E)_{\rm lower~max}<\Delta\x(\E)_{\rm
 max}<\Delta\x(\E)_{\rm upper~max}$, \be\label{range}
 \frac{10.350\Delta\E}{\Delta\x(\E)_{\rm upper~max}}{\rm GeV}^{2}/{\rm fm}<\q<\frac{10.350\Delta\E}{\Delta\x(\E)_{\rm
lower~max}}{\rm GeV}^{2}/{\rm fm} .\ee A representative range of
energies for hard gluons in the QGP produced at RHIC, is $5{\rm
GeV}<E<25{\rm GeV}$. When we make quantitative estimate of $\q$, we
assume the energy of the gluon is 25GeV. The reason is that the
higher the energy is, the longer the penetration length will be, and
the ratio of radiative energy loss to collision energy loss should
be larger, then $\Delta E$ appearing in BDMPS formalism may be
roughly interpreted as the whole energy of the gluon, in short, in
the following estimation, $\Delta E=25{\rm GeV}$. Inserting this
value of $\Delta E$ into Eq. (\ref{range}), we find
\begin{equation}\label{}
    89{\rm GeV}^{2}/{\rm fm}<\hat{q}<106{\rm GeV}^{2}/{\rm fm}.
\end{equation}
This result is much far away from the experimental result  $7{\rm
GeV}^{2}/{\rm fm}<\q<28{\rm GeV}^{2}/{\rm fm}$. The penetration
length in S-S model seems too short, which may be a consequence of
the extra hadronic degrees of freedom.


If we insist that scheme (\ref{scheme}) should be the suitable one
for comparing S-S model plasma with QCD plasma, the good performance
of Gubser's method in $\mathcal{N}$=4 SYM does not take place in S-S
model. So there may be some unknown physical reasons that make
Gubser's method break down when applied to S-S model. It is also
possible that the numeric range of $\hat{q}$ is similar to that of
the experiment may be a coincidence, so we could not require it to
be a feature belonging to all holographic QCD models. But even if
this method can not be always used to estimate $\q$, the relation
between gluon's energy and penetration length may be still
meaningful. However, we do not know how to relate it to experimental
observables.
\section{Appendix}
In this section, we will exhibit a detailed calculation of the
parameter $\zeta$ appearing in Eq. (\ref{scheme}).

We recall that one of the first finite-temperature predictions of
gauge/string duality is that of the thermodynamic potential of dual
gauge theory in the strong coupling regime. The entropy is given by
Bekenstein-Hawking formula $S=A/4G$, where $A$ is the area of the
horizon, $G$ is the ten-dimensional Newton constant. To evaluate
$A$, we cannot use string frame metric (\ref{metric}), but should
use the metric in Einstein frame \cite{Buchel}. The metric in
Einstein frame is obtained from multiplying the string metric
(\ref{metric}) by $\sqrt{g_{s}e^{-\phi}}$, $\phi$ is the dilaton.
The result is \bea \label{einstein-frame} ds^2&=&\left(
\frac{u}{R_{D4}} \right)^{9/8}\left [-f(u) dt^2+
\delta_{ij}dx^{i}dx^j + dx_4^2\right ] +\left( \frac{R_{D4}}{u}
\right)^{15/8} \left [ u^2 d\Omega_4^2 + \frac{du^2}{f(u)} \right ].
 \eea

There are some parameter relations to be used, besides those given
in (\ref{deconfinement})
\begin{eqnarray}
  g_{YM}^{2}&=& 4\pi^{2}g_{s}l_{s}T_{d} \\
  G &=& 8\pi^{6}g_{s}^{2}l_{s}^{8},
\end{eqnarray}
where $T_{S-S}$ is the Hawking temperature of metric (\ref{metric})
and (\ref{einstein-frame}), $T_{d}$ is the critical temperature,
$\lambda$ is the 't~Hooft coupling, $g_{s}$ and $l_{s}$ are string
coupling and string length respectively. Using these relations, we
obtain the entropy density of dual gauge theory by,
\begin{equation}\label{}
    s_{S-S}=\frac{S}{V}=(\frac{2}{3})^{6}\pi^{2}\lambda N_{c}^{2}\frac{T_{S-S}^{5}}{T_{d}^{2}}.
\end{equation}
Applying the following thermodynamic relations between entropy
density $s$, pressure $P$ and free energy density $F$,
\begin{equation}\label{}
    dP=-dF=sdT,
\end{equation}
 we find the pressure and energy density are
\begin{equation}\label{}
    P_{S-S}=\frac{1}{6}Ts_{S-S},
\end{equation}
\begin{equation}\label{}
    \epsilon_{S-S}=\frac{5}{6}Ts_{S-S}.
\end{equation}
It is straightforward to compute the sound speed of S-S plasma
\begin{equation}\label{}
    v_{s}^{2}=\frac{1}{5}.
\end{equation}
This value of sound speed implies that S-S plasma is not a kind of
conformal fluid, for conformal fluid, $v_{s}^{2}$ must be 1/3. We
recall that the energy density of QCD is
\begin{equation}\label{}
    \epsilon_{QCD}=\frac{\pi^{2}}{8}N_{c}^{2}T_{QCD}^{4},
\end{equation}
so we can deduce the relation between $T_{S-S}$ and $T_{QCD}$ by
demanding that $\epsilon_{S-S}=\epsilon_{QCD}$,
\begin{equation}\label{}
    T_{S-S}=T_{QCD}[0.914\lambda^{1/6}(T_{QCD}/T_{d})^{1/3}]^{-1}.
\end{equation}
Then we can extract $\zeta$ from above expression
\begin{equation}\label{}
    \zeta=0.914\lambda^{1/6}(T_{QCD}/T_{d})^{1/3}.
\end{equation}
 This is the Eq. (\ref{zeta}).

\section{Acknowledgements}

I would like to thank M.~Li, Y.~Wang for useful discussions, and I
thank X.~Gao for helping me with typesetting, Y.~Zhou for helping me
revise my paper . Especially, I am grateful to Y.~Wang for helping
me learn using some softwares.


\end{document}